\documentclass{amsart}
\usepackage{cite}
\usepackage{url}
\usepackage[show]{ed}
\usepackage{amsmath}

\def\arccot{\mathop{\rm arccot}\nolimits}
\def\arccoti{{\arccot_1}}
\def\arccotix{{\arccot_9}}
\def\arcsec{\mathop{\rm arcsec}\nolimits}
\def\arccsc{\mathop{\rm arccsc}\nolimits}
\def\arccoth{\mathop{\rm arccoth}\nolimits}
\def\arcsech{\mathop{\rm arcsech}\nolimits}
\def\arccsch{\mathop{\rm arccsch}\nolimits}
\def\arcsinh{\mathop{\rm arcsinh}\nolimits}
\def\arccosh{\mathop{\rm arccosh}\nolimits}
\def\arctanh{\mathop{\rm arctanh}\nolimits}

\def\csch{\mathop{\rm csch}\nolimits}
\def\sech{\mathop{\rm sech}\nolimits}

\def\d{{\rm d}}

\def\ddxi{{\frac{\d^i}{\d x^i}}}

\def\C{{\bf C}}

\def\R{{\bf R}}

\def\myS{{\mathcal S}}

\newtheorem{example}{Example}

\def\INRIA{Algorithms Project\\ Inria Rocquencourt\\ France}
\title{On Kahan's Rules for Determining Branch Cuts}
\author[F. Chyzak]{Fr\'ed\'eric Chyzak}
\address{\INRIA}
\email{Frederic.Chyzak@inria.fr}
\author[J. H. Davenport]{James H. Davenport}
\address{Dept of Computer Science\\ University of Bath\\ United Kingdom}
\email{J.H.Davenport@bath.ac.uk}
\author[C. Koutschan]{Christoph Koutschan}
\address{RISC Linz\\ Johannes Kepler University\\ Austria}
\email{ckoutsch@risc.jku.at}
\author[B. Salvy]{Bruno Salvy}
\address{\INRIA}
\email{Bruno.Salvy@inria.fr}
\thanks{This work was supported in part by the Microsoft Research-INRIA Joint Centre.}

\begin{document}
\maketitle
\begin{abstract}
In computer algebra there are different ways of approaching the mathematical concept of functions, one of which is by defining them as solutions of differential equations. We compare different such approaches and discuss the occurring problems. The main focus is on the question of determining possible branch cuts. We explore the extent to which the treatment of branch cuts can be rendered (more) algorithmic, by adapting Kahan's rules to the differential equation setting.
\end{abstract}
\section{Introduction}
In Mathematics, the standard definition of a \emph{function}
is given, e.g., by Bourbaki~\cite[\S3.4]{Bourbaki1968}:
it consists of the domain and codomain, and requires that a function
be \emph{total} and \emph{single-valued}.
Mathematical practice is usually looser, and tends to define functions locally, or ``in a suitable open subset of $\C$''.
Usually, therefore, users of computer algebra systems will use notations such as ``$\ln(x)$'' without bothering too much about the actual domain and codomain of the function $\ln$ they have in mind, hoping that the designers of the system have implemented what they need. 
An essential and important problem in any such definition is the
determination of possible \emph{branch cuts} (also known as slits),
see~\cite{DingleFateman94}.

A simple way to define a function is by composing previously
defined functions, provided their domains and codomains are
compatible.  This is perhaps best exemplified by
\[
  \sqrt z := \exp\left(\frac12\ln z\right),
\]
which implies that the branch cuts of $\sqrt z$ are inherited
from the definition of~$\ln z$.

The challenge is, of course, to decide which formula we should take. For
instance, as reported in \cite[pp. 210--211]{Kahan1987b}, the classic
handbook \cite{AbramowitzStegun1964} changed its interpretation of the
branch cuts for $\arccot$ from the original first printing (here
denoted as $\arccoti$) to that of the ninth (and subsequent)
printings, and \cite{NIST2010} (here denoted as $\arccotix$).  Both
versions of $\arccot$ are related to $\arctan$ by the formulae:
\begin{align*}
  \arccoti(x) &= \pi/2 - \arctan(x),\\
  \arccotix(x) &= \arctan(1/x).
\end{align*}
The rationale for the ``correct'' choice is discussed further in
Section~\ref{sec:arccot}.

Difficulties occur also when dealing with functions without having
precise definitions for them.  One of the early successes of computer
algebra was symbolic integration. This area is largely based on
differential algebra~\cite{Bronstein2005a}. There, the expression
$\ln(x)$ is not even a function, but some element~$\theta$ in a
differential field, such that $\theta'=1/x$. 
But both $\ln(x)$ and $-\ln(1/x)$ have the same derivative so that
their difference is a constant in the sense of differential algebra:
its derivative is~$0$. However, with the definition of~$\ln$
from~\cite{AbramowitzStegun1964}, the function $\ln(x)+\ln(1/x)$ is~0
for all~$x$ in~$\C$ except for real negative~$x$, where it is~$2i\pi$.
Thus early versions of integrators would wrongly compute
$\int_{-2}^{2}{2x\,dx/(x^2-1)}$ as~0 by first computing correctly the
indefinite integral as $\ln(x^2-1)$ and then subtracting its values at
both end points.

The definition of functions is obviously relevant to identities between functions: the functions involved should have, as a minimum, the same domain and codomain. For instance \cite[pp. 187--188]{Kahan1987b}, the function
\begin{equation}\label{eq:K1}
g(z):= 2\arccosh\left(1+\frac{2z}3\right)-\arccosh\left(\frac{5z+12}{3(z+4)}\right)
\end{equation}
is not the same as the ostensibly more efficient
\begin{equation}\label{eq:K2}
q(z):= 2\arccosh\left(2(z+3)\sqrt{\frac{z+3}{27(z+4)}}\right),
\end{equation}
unless they are defined over an identical domain that avoids the negative real axis and the area
\[   \bigg\{z=x+iy : |y|\le \sqrt{\frac{(x+3)^2(-9-2x)}{2x+5}}\land -9/2\le x\le-3\bigg\}
\]
(and an identical codomain). Clearly, this last statement itself depends on the definitions taken for the functions $\arccosh$ and~$\sqrt{}$. 
Here, the function~$\sqrt{}$ is defined over $\C\setminus{\R^-}$, while $\arccosh$ is defined over~$\C\setminus(1+\R^-)$. In other words, these functions have branch cuts located on horizontal left half-lines. 

In general, the location of branch cuts for classical functions follows very much from the mathematical tradition, as recorded in tables like~\cite{AbramowitzStegun1964}. We discuss here the possibility of automating the choice of these locations for a large class of functions ``defined'' by linear differential equations. We show that this is impossible in general. Therefore, we consider a heuristic approach that gives correct results for the special functions listed in~\cite{AbramowitzStegun1964} and is applicable to newly encountered functions. As a guide, we use Kahan's discussion of this question for inverse trigonometric functions~\cite{Kahan1987b}.

{\rm We are concerned here with automating the choices made, historically by table-makers, and now by the compilers of resources like \cite{NIST2010} or the authors of systems such as \cite{DDMF}, for functions defined by linear differential equations and their inital conditions. A particular context for the {\em use\/} of the functions may impose special constraints, and may even require different choices at different points of the same application \cite[section 4(ii)]{Davenport2010}, but that is a different issue.\/}

\section{Analytic Continuation}

In complex analysis, analytic functions are often defined first on a small domain and then extended by analytic continuation. Two large families of examples are discussed below: inverse functions and solutions of linear differential equations. The basic property underlying this approach is that two analytic functions defined over a connected domain and coinciding over an open subset of it coincide over the whole domain and thus are identical (here we assume the codomain to be~$\C$). Another way of discussing branch cuts is thus in terms of connected domains where the function of interest is to be defined. 

\subsection{Riemann Surfaces}\label{sec:Riemann}
A radical approach is to use \emph{Riemann surfaces} as domains. They
are maximal connected surfaces where the function is analytic.  While
theoretically appealing, the use of Riemann surfaces is not trivial in
a computer algebra context (see, e.g.,~\cite{Hoeven2005}). Since the
domain is not a subset of~$\C$, but paths in the complex place, an
\emph{ad hoc} language for specifying its elements has to be designed.
One possibility is to restrict the domain to paths that are
piecewise-straight lines starting from the origin. This is also the
approach taken in the Dynamic Dictionary of Mathematical
Functions~\cite{DDMF}. Such a path is specified as a list of its
``vertices'', for example, $(0,1+i,2,1-i,0)$ denotes a diamond-shaped,
clockwise path around~$1$. Thus for instance, one can define~$\sqrt{}$
so that it takes the value~$1$ at $(0,1)$, while it is equal to~$-1$
at $(0,1,i,-1,-i,1)$ and to~$1$ again at $(0,1,i,-1,-i,1,i,-1,-i,1)$.

\subsection{Positioning}
In many applications however, users are interested in restricting the domain of their functions to the complex plane or a subset of it. In that case, the role of branch cuts is to define a connected domain where the function is analytic. Where to put the branch cuts is the \emph{positioning  question}.
Apart from the connectivity and analyticity constraints and as long as only one function is involved, the location of the branch cuts is quite arbitrary. The situation is completely different as soon as several functions are involved and identities are considered: the domains have to coincide. Thus branch cuts have to be chosen in a consistent way inside a corpus of functions of interest. 

\subsection{Adherence}
It is customary to extend the domain of definitions to include the branch cuts, so that the function can be defined on the branch cut itself. There, the value of the function is taken as the limit of its values at points approaching the branch cut from one of its sides. In the numerical context, Kahan shows that using signed zeroes avoids having to make a choice~\cite{Kahan1987b}. In the symbolic computation context a choice has to be made and the boundary of the domain is closed on one side and open on the other one. Making the choice of which side is the closed one is the \emph{adherence question} \cite{Beaumontetal2005b}. For instance, the definition of~$\ln$ in~\cite{AbramowitzStegun1964} is taken so that $\ln(-1)$ is~$i\pi$. 
Again, these choices have to be made consistently if several functions are involved.

\subsection{Inverse Functions}
Inverse functions form a large class of functions that are commonly defined by analytic continuation.
Suppose that $f:\C\rightarrow\C$ is analytic, that $f(x_0)=y_0$ and that $f'(x_0)\ne0$. Then there is a trivial function 
\[
  \tilde{f} : \{x_0\}\to\{y_0\},\quad x_0\mapsto y_0
\]
which clearly has an inverse. By the Inverse Function Theorem, this can be extended to a  neighbourhood of $y_0$, and ultimately to the whole of $\C$ apart from those points where $f'(x)=0$. \begin{example}\label{ex:followsqrt2}
The basic example is $\sqrt{}$ defined as the inverse of
\[
  f : \C\to\C,\quad x\mapsto x^2.
\]
Since $f'(1)=2\neq0$, we can define a function~$f^{-1}$ in a neighbourhood of~1 with~$f^{-1}(1)=1$ and then extend it to larger connected domains. However, $f^{-1}$ cannot be continued arbitrarily far round the unit circle, for otherwise we would get the contradicting value $f^{-1}(1)=-1$, see Section~\ref{sec:Riemann}.
\end{example}

\section{Kahan's Rules}
Branch cuts for $\sqrt{z}$, as well as other inverse functions like $\ln z$, $z^\omega$, $\arcsin(z)$, $\arccos(z)$, $\arctan(z)$, $\arcsinh(z)$, $\arccosh(z)$ and $\arctanh(z)$ are given in~\cite{AbramowitzStegun1964}. In all cases, they can be deduced from that of $\ln$ once the function is expressed in terms of~$\ln$ (but even this expression is not neutral, as the second author was initially taught logarithms with a different branch cut). For these functions, Kahan claims~\cite{Kahan1987b}:
\begin{quote}\em There can be no dispute about where to put the slits; their locations are deducible. However, Principal Values have too often been left ambiguous on the slits.
\end{quote}
In the terminology above, this means that the positioning question is soluble, and the problem is the adherence question. He states the following rules governing the location of the branch cuts:
\renewcommand{\labelenumi}{R\arabic{enumi}.}
\begin{enumerate}
\item These functions $f$ are extensions to $\C$ of a real elementary function analytic at every interior point of its domain, which is a segment $\myS$ of the real axis.\label{K:extend}
\item Therefore, to preserve this analyticity (i.e. the convergence of the power series), the slits cannot intersect the interior of $\myS$.\label{K:nonmeet}
\item Since the power series for $f$ has real coefficients, $f(\overline z)=\overline{f(z)}$ in a complex neighbourhood of the segment's interior, so this should extend throughout the range of definition. In particular, complex conjugation should map slits to themselves.\label{K:conjugate}
\item Similarly, the slits of an odd function should be invariant under reflection in the origin, i.e. $z\rightarrow-z$.\label{K:odd}
\item The slits must begin and end at singularities.\label{K:sing}
\end{enumerate}
While these rules are satisfied by the branch cuts of the inverse functions listed above, they do not completely specify their location, unless one adds a form of Occam's razor:
\begin{enumerate}
\item[R6.] The slits might as well be straight lines.\label{K:straight}
\end{enumerate}
We shall interpret R4 in an extended way, by applying it as well when
$f(z) + f(-z)$ is a constant, as will be motivated by the example of
inverse cotangent in Section~\ref{sec:arccot}.

\subsection{Worked example: $\arctan$}\label{K-arctan}
Let us apply these rules to $\arctan$, considered as the inverse of~$\tan$. 
Writing 
\[
  \tan(z)=\frac{e^{iz}-e^{-iz}}{i(e^{iz}+e^{-iz})}
\]
and solving a quadratic equation gives an expression for $\arctan$ in
terms of~$\ln$:
\[
  \arctan(z)=-\frac{i}{2}\Big(\ln(1+ix)-\ln(1-ix)\Big).
\]
From this expression one deduces that the singularities are
located at $\pm i$, so that it is analytic on~$\R$. Moreover, as the
inverse of an odd function, $\arctan$ itself is odd. Hence we need a
cut which
\begin{description}
\item[(R\ref{K:sing})] joins $i$ and $-i$,
\item[(R\ref{K:conjugate})] is invariant under complex conjugation, and
\item[(R\ref{K:odd})] is invariant under $z\rightarrow-z$.
\item[(R6)]We have the choice between a line from $-i$ to $i$ through 0 and two lines $-i-t i$  and $i+t i$, $t>0$, meeting at infinity, but
\item[(R\ref{K:nonmeet})]the first of these two options is not admissible,
\end{description}
giving the classical branch cut $z=0+iy, |y|>1$.

\subsection{The $\arccot$ dilemma}\label{sec:arccot}
The strange case of $\arccot$ described in the introduction is still consistent with these rules. The key point is in~R\ref{K:extend}: in fact $\arccoti$ and $\arccotix$ were defined as different functions over~$\R$: they agreed on $\R^+$ (in particular $\lim_{x\rightarrow+\infty}\arccoti(x)=\lim_{x\rightarrow+\infty}\arccotix(x)=0$), but not on~$\R^-$:
\begin{align*}
  \arccoti(-1) & =3\pi/4,\\
  \arccotix(-1) & =-\pi/4.
\end{align*}
\par
Therefore the limits at~$-\infty$ are different, and in fact $\arccotix$ is continuous at infinity (but discontinuous at 0). What should the branch cuts of these functions be? For $\arccoti$, most of the reasoning of Section~\ref{K-arctan} applies. Strictly speaking, the function is not odd, but it is ``odd apart from a constant'', and hence the branch cuts should still be symmetric under $z\rightarrow-z$. Therefore it should have the same cuts as $\arctan$, i.e. $z=0+iy, |y|>1$.
\par
$\arccotix$ is odd, so all that reasoning applies, except that R\ref{K:nonmeet} no longer rules out the cut passing through 0. Indeed, since $\arccotix$ is discontinuous at $0$, we are left with $z=0+iy, |y|<1$ (the cut in \cite[9th printing]{AbramowitzStegun1964}).

\section{Linear ordinary differential equations}\label{sec:lode}
Many of the elementary, trigonometric, inverse trigonometric functions and hyperbolic versions of those are part of the very large class of solutions of linear differential equations\footnote{Nonlinear equations have the major complication that it may not be obvious where the singularities are, and indeed they may not be finite in number. Some entries of Table~\ref{tab:fns} do not satisfy a linear o.d.e., and this fact is indicated by a dash.} (see Table~\ref{tab:fns}). Here, we set to extend the previous set of rules to fix the location of the branch cuts in a way that is consistent with that of the previous section. Also, we only consider the case where the singularities are all regular singular points (meaning that the solutions have only algebraic-logarithmic behaviour in their neighbourhood).
\begin{table}
\caption{Alternative definitions of functions\label{tab:fns}}
\begin{tabular}{lcc}
Function&Linear o.d.e.&Definition by inverse\\
$\exp$&$y'=y$&$\log^{-1}$\\ 
$\log$&$y'=1/x$&$\exp^{-1}$\\
$\sin$; $\cos$&$y''=-y$&$\arcsin^{-1}$; $\arccos^{-1}$\\
$\tan$; $\cot$&---&$\arctan^{-1}$; $\arccot^{-1}$\\
$\sec$; $\csc$&---&$\arcsec^{-1}$; $\arccsc^{-1}$\\
$\arcsin$; $\arccos$&$y'=\frac{\pm1}{\sqrt{1-x^2}}$&$\sin^{-1}$; $\cos^{-1}$\\
$\arctan$; $\arccot$&$y'=\frac{\pm1}{1+x^2}$&$\tan^{-1}$; $\cot^{-1}$\\
$\arcsec$; $\arccsc$&$y'=\frac{\pm1}{x\sqrt{x^2-1}}$&$\sec^{-1}$; $\csc^{-1}$\\
$\sinh$; $\cosh$&$y''=y$&$\arcsinh^{-1}$; $\arccosh^{-1}$\\
$\tanh$; $\coth$&---&$\arctanh^{-1}$; $\arccoth^{-1}$\\
$\sech$; $\csch$&---&$\arcsech^{-1}$; $\arccsch^{-1}$\\
$\arcsinh$; $\arccosh$&$y'=\frac{1}{\sqrt{x^2\pm1}}$&$\sinh^{-1}$; $\cosh^{-1}$\\
$\arctanh$; $\arccoth$&$y'=\frac{\pm1}{1-x^2}$&$\tanh^{-1}$; $\coth^{-1}$\\
$\arcsech$; $\arccsch$&$y'=\frac{\pm1}{x\sqrt{1\mp x^2}}$&$\sech^{-1}$; $\csch^{-1}$\\
\end{tabular}
\end{table}

Let us assume throughout that we are given a linear ordinary differential equation
\begin{equation}\label{eq:ode}
L(y)=\sum_{i=0}^nc_i \ddxi y=d, \qquad c_i,d \in \C[x].
\end{equation}
At the cost of dividing by $d$, one differentiation and some re-normalisation, we can consider the homogeneous equivalent
\begin{equation}\label{eq:odeh}
L(y)=\sum_{i=0}^{n+1}\hat c_i \ddxi y=0, \qquad \hat c_i \in \C[x].
\end{equation}
More generally, we can homogenize Equation~\eqref{eq:ode} whenever the 
inhomogeneous part $d$~itself satisfies a linear o.d.e.\ of the form~\eqref{eq:odeh}, as is the case with all examples in Table~\ref{tab:fns}.
Outside the zeros of $c_n$ (or $\hat c_{n+1}$), knowing $y$ and sufficiently many of its derivatives at some point $x_0$ (the obvious meaning of \emph{initial conditions}) defines $y$ as an analytic function in the neighbourhood of $x_0$:
\begin{equation}\label{eq:Taylor}
y(x)=y(x_0)+ (x-x_0)y'(x_0)+\cdots,
\end{equation}
where higher derivatives of $y$ beyond the initial conditions are computed by applying \eqref{eq:ode} or~\eqref{eq:odeh} and their derivatives to the initial conditions.
If it weren't for singularities, this would be an excellent definition.

\begin{example}\label{ex:followsqrt}
The function~$\sqrt{}$ can also be defined by
\[
  xy'-\frac12y=0,\quad y(1)=1.
\]
Obviously, the leading coefficient~$x$ has a (regular) singularity at~$0$.
\end{example}

\subsection{Germs of branch cuts}
In the vicinity of a regular singularity, the location of a branch cut can be shown by the form adopted for the local expansion.

For instance, the branch cut for $\arctan$ joins $i$ to $-i$ along the imaginary axis {via infinity} (see Section \ref{K-arctan}). The local behaviour at~$i$ is therefore well described by
\begin{equation}\label{arctanseriesgood}
\arctan(x)= \frac{-i}2\ln(1+ix)+i\ln\sqrt2+\frac14(x-i)+\cdots
\end{equation}
written in such a way that the branch cut ``heads north''.
\par
We can think of the precise formula used to encode the expansion at the singularity as encoding the \emph{germ} of the branch cut, i.e. its local behaviour. The correct angle can always be achieved by rotating the argument.
This solves the positioning problem as far as the germ of the branch cut is concerned. We also need to consider the adherence problem. Eq.~\eqref{arctanseriesgood} inherits the adherence from the logarithm, and therefore, for $y>1$, means that 
\[
  \arctan(0+iy)=\lim_{x\rightarrow0^+}\arctan(x+iy),
\]
which is the adherence described in \cite{Kahan1987b} as ``counter-clockwise continuity''. When we need the other adherence, we simply use
the fact that $\ln(1/x)=-\ln(x)$ except on the branch cut.
\par

\subsection{Heuristic rules}
The adaptation of Kahan's rules to an o.d.e. $L(y)=0$ together with a starting point is as follows: 
\renewcommand{\labelenumi}{R\arabic{enumi}$'$.}
\begin{enumerate}
\setcounter{enumi}{1}
\item The branch cuts do not enter the circle of convergence.
\item Complex conjugation is respected.
\item Any symmetries inherent in the power series are respected.
\item The branch cuts begin and end at singularities.
\item The branch cuts are straight lines.
\item The branch cuts are such that $\C$ less the branch cuts is simply connected.
\end{enumerate}
These subsume Kahan's rules, at the cost of explicitly requiring an initial value, which was implicit in his rules~R\ref{K:extend} and~R\ref{K:nonmeet}. He did not need an equivalent of R7$'$ as his examples only had two singularities. In general, it is required so that the Monodromy Theorem (e.g. \cite[p. 269]{Markushevich1967}) applies and guarantees uniqueness of function values.

These rules \emph{do not} necessarily completely determine the branch cut: a ``random'' differential equation with singularities scattered in the complex plane and no special symmetries will not be determined. Moreover, they do not give any guarantee of consistency between different functions. For instance, both functions~$g$ and~$q$ of~\eqref{eq:K1} and~\eqref{eq:K2} satisfy \emph{the same} linear differential equations. Our rules that lead only to straight lines cannot be compatible with branch cuts that come from compositions of solutions of simpler differential equations with algebraic functions. However, they serve the simple purpose of producing useful and correct branch cuts in a wide variety of cases, including all those discussed before.

\subsection{Worked example: $\arctan$}We apply these rules to $\arctan$, now defined by
\[
  y'=\frac1{1+x^2},\quad y(0)=0.
\]
The singularities of this differential equation are clearly at $x=\pm i$, and the function so defined is odd.  Hence we need a cut which:
\begin{description}
\item[(R\ref{K:sing}$'$)]joins $i$ and $-i$,
\item[(R6$'$)]does it in a straight line,
\item[(R\ref{K:conjugate}$'$)]is invariant under complex conjugation,
\item[(R\ref{K:odd}$'$)]is invariant under $z\rightarrow-z$,
\item[(R2$'$)]does not enter the unit disk.
\end{description}
Thus we find again the classical branch cut $z=0+iy, |y|>1$. We then deduce expansions at the singularities that match the germs of this cut as in~(\ref{arctanseriesgood}).

\subsection{$5\ln x$ or $\ln x^5$?}
Once $\ln$ has been defined, the functions~$F_1=5\ln x$ and $F_2=\ln x^5$ are different:
$F_1(i)=\frac{5\pi i}2$ while $F_2(i)=\frac{\pi i}2$. Nevertheless, they are both solutions to $xy'-5=0$ with $y(1)=0$. Our approach would make the choice $5\ln x$ with only one branch cut, while~$F_2$ has five branch cuts, at angles of $\{1,3,5,7,9\}\pi/10$, thus making the domain not connected (and violating R7$'$).

\subsection{A harder example}\label{sec:arctanz2}
Let us consider the functions $f$ defined by
\[
f'=\frac{2x}{1+x^4} , \]
or, if one prefers homogeneous equations,
\[
x(1+x^4)f''+(3x^4-1)f'=0.\]
In both cases, we assume we are given real initial conditions at~0.
This example is selected because it has two simple, linearly independent solutions---1 and~$\arctan(x^2)$---to compare with the result, but the method does not use this information and would apply even if no such solution could be found. So here is what we get:
\begin{description}
\item[(R\ref{K:sing}$'$)] The equation has four regular singularities at
\[
  z=\pm\sqrt{\pm i}=\frac{\pm1\pm i}{\sqrt 2}
\]
(one can check that~$0$ is just an apparent singularity by exhibiting
a basis of formal power series solutions and that $\infty$ is not a
singularity by changing~$x$ into~$1/x$).
\item[(R6$'$)] These four singularities have to be connected by
  straight lines.
\item[(R\ref{K:nonmeet}$'$)] We cannot connect the singularities
  pairwise (in either way!) without going to infinity.
\item[(R\ref{K:odd}$'$)] The symmetry $f(ix)=-f(x)$ can be checked
  directly from the equation, so that branch cuts should be mapped to
  branch cuts by a rotation of~$\pi/2$. 
\item[(R\ref{K:conjugate}$'$)] Reality implies
  that branch cuts are also mapped to branch cuts by horizontal
  symmetry. 
\end{description}
We are thus left with only the following choice:
Cuts that ``head northeast'' from $\frac{1+i}{\sqrt 2}$, ``northwest'' from $\frac{-1+i}{\sqrt 2}$ etc., all meeting at infinity. This is indeed consistent with~$\arctan(x^2)$.
\par
It is worth noting that this function actually also admits branch cuts that violate R\ref{K:nonmeet}$'$ and R7': for example we can connect $\frac{-1- i}{\sqrt 2}$ to $\frac{+1- i}{\sqrt 2}$, and $\frac{-1+ i}{\sqrt 2}$ to $\frac{+1+ i}{\sqrt 2}$. This is a peculiarity of our construction, and the fact that these are valid follows, not from the Monodromy Theorem, but from the fact that the residues at these branch points are equal and opposite.

\section{Conclusions}
When it comes to converting an analytic (be it linear ordinary differential equation, inverse function, or possibly other) definition of a function into a well-defined single-valued one, so that one can answer questions such as ``what is $\ln(-1)$?'' or ``what is $\arctan(2i)$'', branch cuts may need to be imposed on the locally analytic function. While the definition of the function may stipulate the endpoints of the cut, it does not, in general, specify the location of the cut between its endpoints, nor indeed even the germ of the cut at the singularities.

We have given a simple set of rules that is convenient when nothing else is known about the function. This set of rules is sufficient to recover the classical branch cuts of the elementary inverse trigonometric or hyperbolic trigonometric functions. However, it is important to remember that this is only a useful heuristic, while there are cases where different cuts are dictated by the application.
In a specific context, getting the right overall function defined by a
formula can even require \emph{inconsistent\/} choices of the branch
cuts of component functions: see, e.g., the Joukowski map studied
in~\cite[pp. 294--8]{Henrici1974}) and reported on
in~\cite{Davenport2010}.

\par
{\bf Acknowledgements.} The authors are grateful to other members of the Algorithms team, notably Alexandre Benoit, Marc Mezzarobba and Flavia Stan, for their contributions. The second author is grateful to Peter Olver for 
some clarifying discussions on the topic.

\end{document}